\documentclass[sigconf, nonacm]{acmart}

\usepackage{amsmath}
\usepackage{xspace}
\usepackage{enumitem}
\setlist[itemize]{leftmargin=*}
\setlist[enumerate]{leftmargin=*}
\usepackage{graphicx}
\usepackage[caption=false]{subfig}
\usepackage{cleveref}
\usepackage{pgfplots}
\usepackage{pgfplotstable}
\pgfplotsset{compat=1.18}
\usepgfplotslibrary{groupplots}
\usepackage[most]{tcolorbox}
\usepackage{pifont}
\usepackage{balance}

\definecolor{tidehunterblue}{RGB}{31,119,180}
\definecolor{rocksdborange}{RGB}{255,127,14}
\definecolor{blobdbgreen}{RGB}{44,160,44}
\definecolor{headerblue}{RGB}{31,119,180}
\definecolor{optimisticorange}{RGB}{255,127,14}

\pgfplotsset{
    barplot/.style={
            ybar,
            bar width=0.25cm,
            enlarge x limits=0.25,
            ylabel={Throughput (Kops/s)},
            xlabel={Read Percentage (\%)},
            symbolic x coords={0,50,100},
            xtick=data,
            ymin=0,
            enlarge y limits={upper,value=0.18},
            legend style={at={(0.98,0.98)}, anchor=north east, font=\tiny, cells={anchor=west}},
            nodes near coords,
            every node near coord/.append style={font=\tiny, rotate=90, anchor=west},
            width=0.32\textwidth,
            height=0.20\textheight,
            tick label style={font=\small},
            label style={font=\small},
        }
}

\pgfplotsset{
    widebarplot/.style={
            ybar,
            bar width=0.12cm,
            enlarge x limits=0.06,
            ylabel={Throughput (Kops/s)},
            symbolic x coords={100\% write,50\% get,50\% exists,50\% lt,100\% get,100\% exists,100\% lt},
            xtick=data,
            x tick label style={font=\scriptsize, rotate=45, anchor=east},
            ymin=0,
            enlarge y limits={upper,value=0.18},
            legend style={at={(0.02,0.98)}, anchor=north west, font=\scriptsize, cells={anchor=west}},
            nodes near coords,
            every node near coord/.append style={font=\tiny, rotate=90, anchor=west},
            width=\columnwidth,
            height=0.20\textheight,
            tick label style={font=\scriptsize},
            label style={font=\scriptsize},
        }
}

\pgfplotsset{
    indexlineplot/.style={
            width=0.48\textwidth,
            height=0.25\textheight,
            xlabel={Thread Count},
            ylabel={Throughput (Kops/s)},
            xmin=0, xmax=50,
            ymin=0,
            grid=major,
            grid style={gray!30},
            legend style={at={(0.02,0.98)}, anchor=north west, font=\tiny, cells={anchor=west}},
            tick label style={font=\small},
            label style={font=\small},
            mark size=2pt,
            line width=1pt,
        },
    indexwindowplot/.style={
            width=0.325\textwidth,
            height=0.18\textheight,
            xlabel={Window Size},
            ylabel={Throughput (Kops/s)},
            xmin=80, xmax=4000,
            xtick={100,200,400,800,1600,3200},
            xticklabels={100,200,400,800,1600,3200},
            ymin=0,
            grid=major,
            grid style={gray!30},
            legend style={at={(0.02,0.98)}, anchor=north west, font=\scriptsize, cells={anchor=west}},
            tick label style={font=\scriptsize},
            label style={font=\scriptsize},
            mark size=1.5pt,
            line width=0.8pt,
        },
    indexwindowplot nolabel/.style={
            indexwindowplot,
            ylabel={},
            trim axis left,
        }
}

\newif\iffullversion
\fullversiontrue 

\newcommand{\sysname}{\textsc{Tidehunter}\xspace}

\tcbset{
    highlight/.style={
            enhanced,
            breakable,
            colback=gray!10,
            frame hidden,
            boxrule=0pt,
            borderline west={2pt}{0pt}{black},
            left=2mm, right=2mm, top=1mm, bottom=1mm,
        }
}

\newcounter{takeaway}

\newcounter{insight}

\newcommand{\one}{\ding{182}\xspace}
\newcommand{\two}{\ding{183}\xspace}
\newcommand{\three}{\ding{184}\xspace}
\newcommand{\four}{\ding{185}\xspace}
\newcommand{\five}{\ding{186}\xspace}

\iffullversion\else
\newcommand\vldbdoi{XX.XX/XXX.XX}
\newcommand\vldbpages{XXX-XXX}
\newcommand\vldbvolume{14}
\newcommand\vldbissue{1}
\newcommand\vldbyear{2020}

\newcommand\vldbauthors{\authors}
\newcommand\vldbtitle{\shorttitle}

\newcommand\vldbavailabilityurl{https://github.com/MystenLabs/tidehunter}

\newcommand\vldbpagestyle{plain}
\fi 

\begin{document}

\date{}

\title{
  \sysname: Large-Value Storage With Minimal Data Relocation
}

\author{Andrey Chursin}
\affiliation{\institution{Mysten Labs}}
\email{andrey@mystenlabs.com}

\author{Lefteris Kokoris-Kogias}
\affiliation{\institution{Mysten Labs}}
\email{lefteris@mystenlabs.com}

\author{Alex Orlov}
\affiliation{\institution{Mysten Labs}}
\email{alexo@mystenlabs.com}

\author{Alberto Sonnino}
\affiliation{\institution{Mysten Labs and UCL}}
\email{alberto@mystenlabs.com}

\author{Igor Zablotchi}
\affiliation{\institution{Mysten Labs}}
\email{igor@mystenlabs.com}

\begin{abstract}
  Log-Structured Merge-Trees (LSM-trees) dominate persistent key-value storage but suffer from high write amplification from 10$\times$ to 30$\times$ under random workloads due to repeated compaction. This overhead becomes prohibitive for large values with uniformly distributed keys, a workload common in content-addressable storage, deduplication systems, and blockchain validators. We present \sysname, a storage engine that eliminates value compaction by treating the Write-Ahead Log (WAL) as permanent storage rather than a temporary recovery buffer. Values are never overwritten; and small, lazily-flushed index tables map keys to WAL positions. \sysname introduces (a) lock-free writes that saturate NVMe drives through atomic allocation and parallel copying, (b) an optimistic index structure that exploits uniform key distributions for single-roundtrip lookups, and (c) epoch-based pruning that reclaims space without blocking writes. On a 1\,TB dataset with 1\,KB values, \sysname achieves 830K writes per second, that is 8.4$\times$ higher than RocksDB and 2.9$\times$ higher than BlobDB, while improving point queries by 1.7$\times$ and existence checks by 15.6$\times$.  We validate real-world impact by integrating \sysname into Sui, a high-throughput blockchain, where it maintains stable throughput and latency under loads that cause RocksDB-backed validators to collapse. \sysname is production-ready and is being deployed in production within Sui.
\end{abstract}

\maketitle

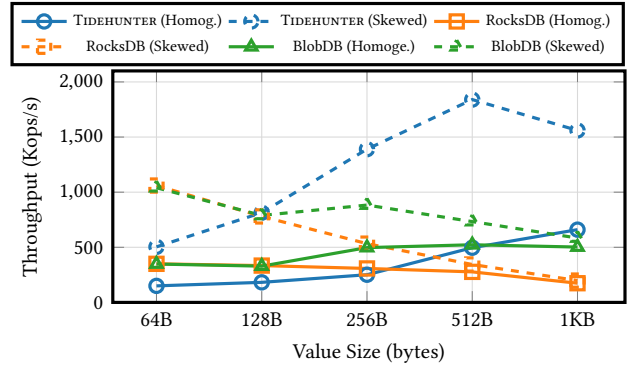
\begin{figure}[t]
    \centering
    \begin{tikzpicture}
        \begin{axis}[
                width=0.99\columnwidth,
                height=0.55\columnwidth,
                xlabel={Value Size (bytes)},
                ylabel={Throughput (Kops/s)},
                xmode=log,
                log basis x=2,
                xmin=48, xmax=1400,
                xtick={64,128,256,512,1024},
                xticklabels={64B,128B,256B,512B,1KB},
                ymin=0,
                ymax=2100,
                grid=major,
                grid style={gray!30},
                legend style={at={(0.4,1.02)}, anchor=south, font=\scriptsize, legend columns=3},
                tick label style={font=\small},
                label style={font=\small},
                mark size=2.5pt,
                line width=1.2pt,
            ]
            \addplot[color=tidehunterblue, mark=o, solid] coordinates {
                    (64,150) (128,182) (256,251) (512,496) (1024,660)
                };
            \addplot[color=tidehunterblue, mark=o, dashed] coordinates {
                    (64,504) (128,814) (256,1390) (512,1840) (1024,1560)
                };
            \addplot[color=rocksdborange, mark=square, solid] coordinates {
                    (64,350) (128,334) (256,308) (512,277) (1024,174)
                };
            \addplot[color=rocksdborange, mark=square, dashed] coordinates {
                    (64,1060) (128,780) (256,533) (512,344) (1024,198)
                };
            \addplot[color=blobdbgreen, mark=triangle, solid] coordinates {
                    (64,348) (128,329) (256,497) (512,523) (1024,502)
                };
            \addplot[color=blobdbgreen, mark=triangle, dashed] coordinates {
                    (64,1040) (128,789) (256,883) (512,733) (1024,584)
                };
            \legend{\sysname{} (Homog.), \sysname{} (Skewed), RocksDB (Homog.), RocksDB (Skewed), BlobDB (Homoge.), BlobDB (Skewed)}
        \end{axis}
    \end{tikzpicture}
    \vspace{-0.5cm}
    \caption{\footnotesize Mixed workload throughput (50\% reads/writes) vs.\ value size, with skewed (Zipf $\theta=2$) and homogeneous (Zipf $\theta=0$) access patterns.}
    \vspace{-0.5cm}
    \label{fig:value-scaling}
\end{figure}

\section{Introduction}\label{sec:intro}



Log-Structured Merge-Trees (LSM-trees) are the dominant design for persistent
key-value stores. An LSM-tree buffers writes in memory and flushes them to
disk as sorted files. To bound the number of files and maintain lookup
performance, the storage engine periodically merges these files in a process
called compaction. Compaction rewrites data to maintain sorted order. For
small values, the overhead is acceptable. For values of several kilobytes, it
is not. A common measure of this inefficiency is write amplification, i.e.,
bytes written to disk divided by bytes received from the application. Write
amplification reaches $10\times$ to $30\times$~\cite{rocksdb2021evolution} under random workloads and
grows with dataset size. A store ingesting 100\,MB/s of application data may
push 1\,GB/s to disk, starving reads of bandwidth.

LSM-trees target hard disk drives (HDDs), where sequential I/O is $100\times$
faster than random, hence trading write amplification for sequential access was the recommended approach. Solid-state drives (SSDs) changed the equation. Sequential-to-random
gaps shrink to $10\times$ or less for large requests, and SSDs expose internal
parallelism that LSM-trees cannot exploit. To make matters worse,
SSD cells also wear out after a finite number of writes, making high write amplification the culprit of frequent device failures.

A seminal work to address this issue is WiscKey~\cite{lu2016wisckey}.
It introduced key-value separation: keys stay in an
LSM-tree, values go to a separate append-only log. Values stop moving during
compaction, and write amplification drops.  However, the LSM-tree remains,
and so do its compaction cost and multi-level lookup overhead. Under
uniform key distributions, lookups traverse multiple levels with little cache
benefit. Additionally, the value log of WiscKey needs garbage collection, which stalls I/O.

We take a different path. In \sysname, the Write-Ahead Log (WAL), traditionally a
temporary buffer for crash recovery, becomes the permanent store for values.
We replace the key LSM-tree with sharded index tables that flush lazily and
answer existence queries from memory. Our design targets applications with large values and uniformly distributed keys.
These are common requirements of modern applications.
Content-addressable storage systems like Git and
IPFS key objects by cryptographic hashes of their
contents~\cite{ipfs2014}. Deduplication systems for backup and archival
storage partition data into chunks and index them by hash, with chunk sizes
typically ranging from 4\,KB to 64\,KB~\cite{zhu2008avoiding,
      guo2011building}. Object stores and blob storage systems often use UUIDs or
content hashes as keys; Facebook's Haystack, for example, stores billions of
photos keyed by unique identifiers with a size of around
8-64\,KB~\cite{beaver2010finding}. Machine learning feature stores index
embeddings and feature vectors by entity ID or feature hash, with vectors
often reaching several kilobytes~\cite{hopsworks2024,devlin2019bert,orr2021managing}. The shift toward
immutable, content-addressed, and UUID-keyed data spans cloud infrastructure,
data pipelines, and application backends. In all these settings, keys lack
locality by design, and values are large enough that LSM-tree compaction
becomes the dominant cost.

We focus on blockchain validators as a concrete evaluation target because they
combine all of these pressures in a single system: hash-keyed data, kilobyte
values, write-heavy ingestion, latency-sensitive reads, and aggressive pruning.
A validator participates in distributed consensus: it receives transactions,
orders them into blocks, executes them, and stores results for queries. We
specifically chose Sui~\cite{blackshear2024sui} to integrate \sysname. It is a
high-throughput blockchain whose validators handle thousands of transactions per
second and answer queries about history, execution results, and object state
concurrently. Validators write continuously as blocks arrive but must read with
low latency, since execution fetches objects, clients check status, and peers
request proofs. They also prune aggressively, keeping only recent epochs; this
places significant stress on the underlying RocksDB store, where ill-timed
compactions can cause performance to plummet.

\sysname is built for large values and uniform keys. \Cref{fig:value-scaling} shows the performance benefit of \sysname as the value size grows, both for skewed and homogeneous access patterns. Three key insights guided our design towards these results:

\begin{enumerate}
      \item \textbf{Separate values from indices.} Values live in a central WAL;
            indices are small and separate. Compaction rewrites indices, not values.
            Large values are written once and never copied.
      \item \textbf{Manage indices lazily.} Indices load on-demand and unload
            under memory pressure, enabling working sets larger than RAM without
            manual tuning.
      \item \textbf{Exploit epoch semantics for garbage collection.} Many
            workloads, including blockchains, organize data into epochs or time
            ranges. \sysname reclaims space by dropping entire WAL segments when their
            epoch expires, avoiding the I/O cost of per-record garbage collection.
\end{enumerate}

We evaluate \sysname against RocksDB and BlobDB (the production implementation of WiscKey) on 1\,TB datasets. At 1\,KB
values, \sysname sustains 830K writes/sec ($8.4\times$ RocksDB, $2.9\times$
BlobDB), improves point queries by $1.7\times$, and improves existence checks
by $15.6\times$ (resolved from the index, no value fetch). At 64 bytes,
LSM-trees win by $2$--$2.5\times$; the crossover is near 128 bytes.

To validate that synthetic benchmarks translate to real systems, we integrated
\sysname into Sui and ran validators under sustained transaction load. With
RocksDB, validators show performance degradation at around 6,000 transactions
per second as disk utilization rises and latency climbs. Under the same
conditions, \sysname maintains stable throughput and lower latency, with
reduced disk pressure and more consistent CPU utilization. Side-by-side runs
show clear differences in disk read and write throughput, sustained
transactions per second, and settlement finality latency. \sysname avoids the
I/O saturation that triggers performance collapse in RocksDB configurations at
comparable load.

\paragraph{Contributions} This paper makes the following contributions:
\begin{itemize}
      \item We present \sysname, a storage engine architecture that eliminates
            value compaction by treating the WAL as permanent storage, achieving near
            $1\times$ write amplification for large values.
      \item We introduce an optimistic index that exploits uniform key
            distributions to achieve single-roundtrip lookups, trading the generality of LSM-trees for lower tail latencies.
      \item We evaluate \sysname on synthetic benchmarks and a production
            blockchain, showing $8.4\times$ write throughput over RocksDB and stable
            performance where RocksDB-backed systems collapse.
\end{itemize}

\section{Background \& Design Principles}\label{sec:overview}

We begin by reviewing LSM-trees and their compaction costs, which motivate \sysname's design. We then present the high-level principles that guide \sysname's architecture.

\begin{figure*}[t]
    \centering
    \includegraphics[width=0.9\textwidth]{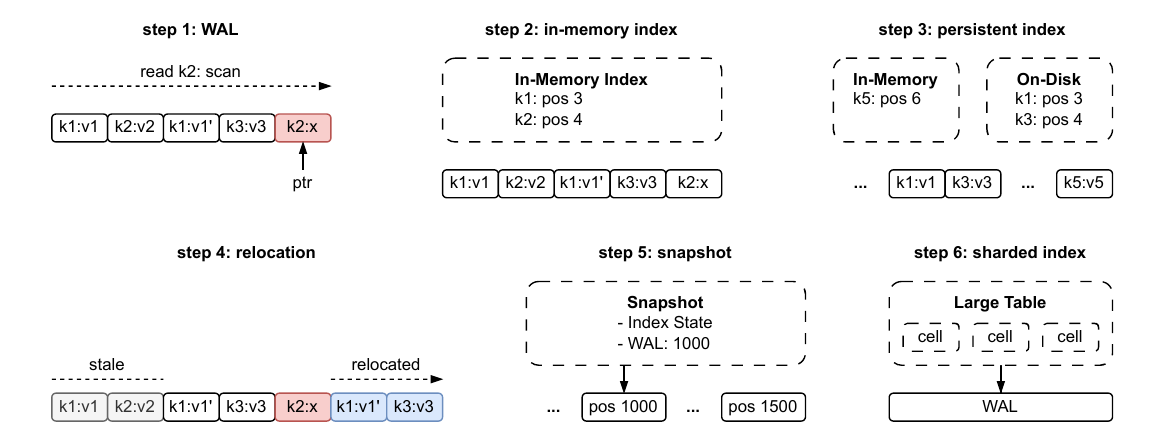}
    \Description{\footnotesize Six-step evolution of TideHunter's design: Step 1 append-only log, Step 2 in-memory index, Step 3 persisted index, Step 4 relocation, Step 5 snapshots, Step 6 sharded index.}
    \caption{\footnotesize Design evolution of \sysname. Each step addresses a limitation of the previous design: (1) append-only log provides fast writes but slow reads; (2) in-memory index enables fast lookups; (3) persisted index supports datasets larger than memory; (4) relocation reclaims space without blocking writes; (5) snapshots enable fast recovery; (6) sharding enables concurrent operations.}
    \label{fig:overview}
    \vskip -1em
\end{figure*}

\subsection{Log-Structured Merge-Trees}

LSM-trees~\cite{oneil1996lsm} emerged as the
dominant architecture for write-intensive key-value stores. The design reflects
a fundamental asymmetry of hard disk drives: sequential I/O is roughly
$100\times$ faster than random I/O~\cite{rosenblum1992lfs,he2023storage}. Rather than update
records in place, which
requires random seeks, LSM-trees buffer writes in memory and flush them
sequentially. This batching converts random writes into sequential at the
cost of extra bookkeeping.

An LSM-tree organizes data into levels. Incoming writes first land in
\emph{memtable}, an in-memory sorted structure such as a balanced tree or
skiplist. When the memtable reaches a size threshold, the engine flushes it to
disk as an immutable Sorted String Table (SST) file. These flushed files form
Level~0 (L0). Unlike deeper levels, L0 files may have overlapping key ranges
because each flush produces an independent file.
As L0 accumulates files, read performance degrades: a lookup checks the
memtable, then each L0 file from newest to oldest, because any of them might
contain the key. To bound this, the engine periodically runs
\emph{compaction}, which merges L0 files into Level~1 (L1). L1 and deeper
levels maintain a key invariant: within a level, files have non-overlapping key
ranges. A lookup at L1 or beyond touches aa single file per level. Compaction
continues down the tree: when L1 grows too large, files merge into L2, and so
on.

RocksDB~\cite{rocksdb2021evolution}, a widely deployed LSM-tree
implementation, uses this leveled compaction strategy by default. Each level is roughly $10\times$ larger than
the previous, so a dataset of size $N$ requires $O(\log N)$ levels. Compaction
is crucial for read performance: without it, every lookup would scan the entire
L0, making reads linear in the number of flushes.

\subsection{Compaction Costs}

Compaction maintains read performance but introduces three forms of overhead
that dominate under write-heavy workloads.

\paragraph{Write amplification.}
Each key-value pair may be rewritten many times as it moves through levels.
When compaction merges files from L$i$ into L$i+1$, every record in those files
is read and written again, even if it has not changed. With a size ratio of
$10\times$ between levels and $O(\log N)$ levels, write amplification---bytes
written to disk divided by bytes received from the application---can reach
$10\times$ to $30\times$ under random workloads~\cite{rocksdb2021evolution}. In Sui's private testnet,
validators ingesting 40\,MB/s of application data generated 500\,MB/s of disk
writes, an $8\times$ amplification factor. High write amplification consumes
I/O bandwidth, accelerates SSD wear~\cite{hu2009writeamp}, and limits throughput
scaling.

\paragraph{Read amplification.}
Although compaction reduces the number of files a lookup must check, reads
still traverse multiple levels. A point query may touch the memtable, one or
more L0 files, and one file per deeper level. Bloom filters mitigate this cost
for negative lookups~\cite{rocksdb2021evolution}, but positive lookups on cold
data incur one disk read per level. For a 1\,TB dataset with $10\times$ level ratios, this means five or six
levels and potentially five or six disk reads per lookup.

\paragraph{Write stalls.}
Compaction competes with foreground writes for I/O bandwidth and CPU cycles. To
prevent L0 from growing without bound, RocksDB throttles or blocks writes when
compaction falls behind. These \emph{write stalls} cause latency spikes and
throughput drops~\cite{rocksdb2021evolution, balmau2019silk}. In extreme cases, writes block entirely until compaction
clears enough space. The result is unpredictable performance: the same workload
may sustain high throughput one minute and stall the next.
These costs compound as the dataset grows and values become larger. In Sui's
testnet, disk I/O utilization approached 80\%, leaving no headroom to scale
throughput. The bottleneck was not the application---it was the storage engine
rewriting data that had not changed. \sysname addresses this by eliminating
value compaction entirely: values land in the WAL once and never move again.

\subsection{\sysname Design Principles} \label{sec:principles}

The limitations of LSM-Tree-based designs specifically for applications with large values and uniform keys has guided our design for \sysname.
\Cref{fig:overview} presents our journey towards \sysname by starting from a minimal storage system and incrementally addressing its limitations. Each step introduces a component that appears in the final architecture (\Cref{fig:overview}). Three principles guide every decision: (1) writes append directly to their long-term position on disk, (2) values are rarely moved, and (3) compaction never blocks writes.

\paragraph{Step 1: Append-Only Log}
The simplest design stores all data in a single append-only log. Inserts and updates append the key-value pair to the end of the log. Deletes append a tombstone marker. Reads perform a linear scan from the end of the log, returning the first matching key encountered.
This design achieves $O(1)$ writes but $O(n)$ reads. This is acceptable for write-heavy workloads with few reads, but impractical otherwise. We need an index.

\paragraph{Step 2: In-Memory Index}
We add an in-memory index that maps keys to their positions in the log. Writes still append to the log, but now also update the index. Reads consult the index to find the log position, then fetch the value directly---$O(\log n)$ or $O(1)$ depending on the index structure.
However, the index grows with the number of unique keys. For datasets larger than memory, the index itself becomes the bottleneck. We need to persist it.

\paragraph{Step 3: Persisted Index}
We periodically flush the in-memory index to disk. The flushed index is a sorted file that can be searched efficiently. Reads first check the in-memory index for recent writes, then fall back to the persisted index on disk.
Now the system can handle datasets larger than the RAM. But the WAL grows unbounded: deleted keys and overwritten values consume space forever. We need to reclaim it.

\paragraph{Step 4: Relocation}
We introduce a relocation process that moves live entries forward in the log and discards obsolete ones. The relocator scans from the oldest entries, checks whether each key is still live (not deleted or overwritten), and re-appends live entries to the log tail. Once all live entries in a log segment have been relocated, the segment can be deleted.
Crucially, relocation runs in the background and never blocks writes. Unlike LSM-tree compaction, which must complete before the system can proceed, \sysname's relocator simply appends to the same log that serves normal writes.
However, recovery currently requires replaying the entire WAL to reconstruct the index, which is impractically slow for large datasets.

\paragraph{Step 5: Snapshots}
We add periodic snapshots that record the index state and a WAL position on disk. Recovery loads the snapshot and replays only the WAL suffix after the snapshot position.
Snapshots are small because they store only metadata (index positions and WAL offsets), not the actual index data. They can be taken frequently without significant overhead.
Finally, a single index becomes a contention point under concurrent writes: locking it even for a brief time for updates can severely hinder performance.

\paragraph{Step 6: Sharded Index}
We partition the key space into independent cells, each with its own in-memory index, persisted index, and locks. Operations on different shards proceed in parallel without synchronization.
This is \sysname in a nutshell. The WAL remains shared (writes are serialized only at allocation), but index operations parallelize across shards. \Cref{sec:design} describes each component and the challenges that this design overcomes in detail.

\section{The \sysname Architecture}\label{sec:design}

\begin{figure}[t]
    \centering
    \includegraphics[width=0.8\columnwidth]{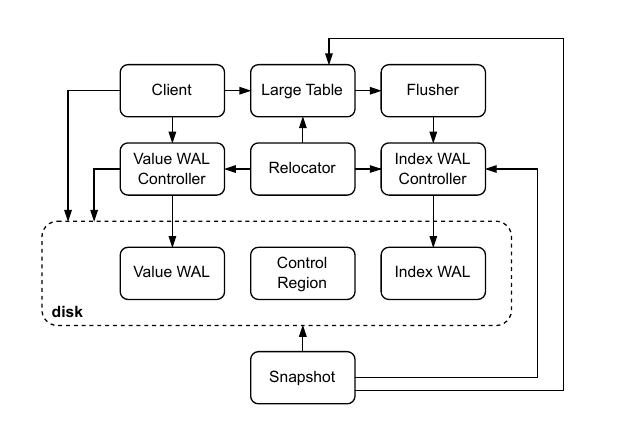}
    \Description{Architecture diagram showing the main components of Tidehunter: Client, Large Table with LRU cache and Bloom filter, Value WAL with its Controller, Index Store with its Controller, Flusher, Snapshot engine writing to Control Region, and Relocator. Arrows show data flow between components.}
    \vskip -1em
    \caption{High-level architecture of \sysname.}
    \label{fig:architecture}
    \vskip -1em
\end{figure}

\Cref{fig:architecture} illustrates the architecture.
The system consists of several interacting components.
This section describes the interaction between these components to serve reads and writes as well as ensuring efficient crash-recovery.
%

\subsection{Write Flow} \label{sec:write}

\begin{figure}[t]
    \centering
    \includegraphics[width=0.8\columnwidth]{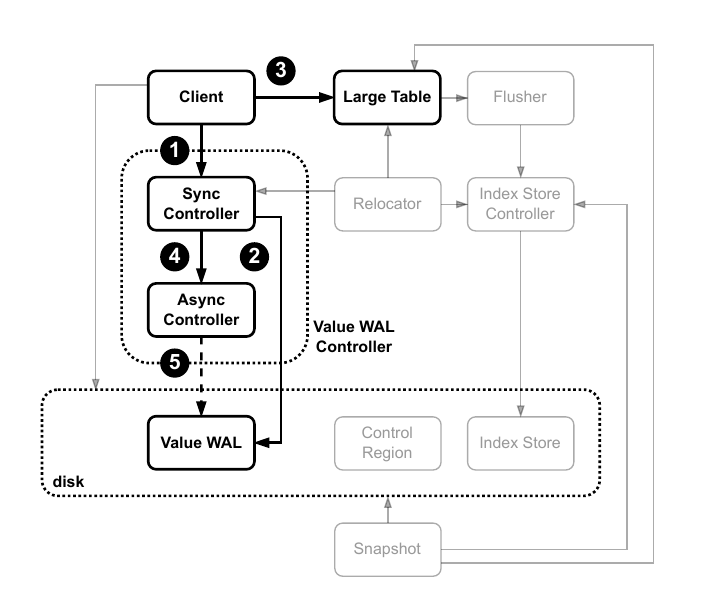}
    \Description{Write path diagram showing the synchronous and asynchronous components. Solid arrows show the synchronous path: client to synchronous controller for allocation, memory-map write to Value WAL, and update to Large Table. Dotted arrows show asynchronous operations: notification to asynchronous controller, which manages map lifecycle and calls fsync.}
    \vskip -1em
    \caption{Write path separating fast synchronous operations (solid) from background durability (dotted).}
    \label{fig:write}
    \vskip -1em
\end{figure}

\Cref{fig:write} details the write path by expanding the Value WAL Controller of \Cref{fig:architecture}. The write path is designed to minimize latency while providing clear durability guarantees. It contains a synchronous path that completes fast (solid arrows on \Cref{fig:write}), and asynchronous background operations (dotted arrows on \Cref{fig:write}).

\paragraph{The Client flow}
The client starts by sending a write operation (insert or delete) to the synchronous controller, which allocates a position in the Value WAL, that is, a contiguous region where the entry will be written, to ensure concurrent writers do not overwrite each other~(\one). Its role is to monitor which WAL positions have been fully processed (i.e., both written to the WAL and indexed in the Large Table). It then memory-map writes the entry to the Value WAL, persisting it to disk~(\two).
The client then updates the Large Table, the in-memory index that maps keys to their WAL positions, so that subsequent reads can locate the value directly without scanning the entire WAL~(\three).
This allows multiple clients to start the write flow concurrently. The synchronous controller allocates WAL positions through a fast atomic counter and after that, the clients run the rest of the write flow in parallel.

When the update completes, the asynchronous controller is notified that this position is now fully processed, meaning the entry is both durable in the WAL and indexed in the Large Table~(\four). It updates the highest WAL offset for which all preceding entries have been processed. This position, along with a snapshot of the Large Table, is periodically persisted to the Control Region to facilitate efficient crash recovery (see \Cref{sec:recovery}).

The Value WAL is organized into memory-mapped regions called maps, each covering a contiguous range of positions. Multiple entries are written to the same map before it is finalized. When all positions in a map have been processed and the next map has begun receiving writes, the asynchronous controller asynchronously finalizes the map.
The asynchronous controller serves three purposes: it manages the lifecycle of these memory maps, handles garbage collection of old WAL files, and calls fsync to finalize maps. For map lifecycle, it maintains a buffer of pre-allocated maps so that writes never block waiting for map creation. When a map is finalized, it prepares the next map in the buffer; it then asynchronous calls fsync on finalized maps, ensuring data is persisted to stable storage~(\five). By isolating this blocking operation in background threads, the write path remains fast and non-blocking. For garbage collection, when relocation advances the minimum required WAL position, the asynchronous controller deletes obsolete WAL files that are no longer needed for recovery (see \Cref{sec:relocation}).

\paragraph{Durability Guarantees}
This design cleanly separates the synchronous write path: allocation, memory-map write, and index update, from the asynchronous durability and cleanup operations. Writes complete without waiting for fsync, while background threads ensure data eventually reaches stable storage and stale files are reclaimed.
\sysname provides durability against application crashes as soon as step \three completes, when data is written to the memory-mapped region. Even if the application crashes before the Large Table is updated, the kernel will eventually flush the page cache to disk. Upon recovery, \sysname replays the WAL from the last checkpointed position, reconstructing any index entries that were not yet reflected in the Large Table. For durability against kernel crashes, \sysname relies on explicit fsync operations performed asynchronously by the background asynchronous controller. Applications requiring stronger durability guarantees can invoke explicit flush operations.
Having the synchronous controller separated from the asynchronous controller allows for a multi-tier performance/durability trade-off. The former finishes fast and provides persistence at the OS page cache, while the latter batches fsync calls and handles cleanup in the background.

\paragraph{Deletes}
Deleting a key is analogous to writing a special tombstone value. If concurrent operations occur (e.g., a delete and an insert for the same key), the operation with the higher WAL position wins. This ensures consistency regardless of the order in which operations are applied to the Large Table. Tombstones serve three purposes: they ensure deletes survive crash recovery (replaying the tombstone removes the key), they allow correct conflict resolution with concurrent writes, and they enable the relocator to skip deleted entries when compacting the Value WAL (\Cref{sec:relocation}).

\paragraph{Atomic batch writes}
\sysname supports atomic batch writes, allowing multiple inserts and deletes to be applied as a single unit. A batch is constructed by accumulating updates in memory, then committed in two phases. First, a batch-start marker is written to the Value WAL, followed by all entries in the batch. The Value WAL Controller uses a single allocation for the entire batch, ensuring they are treated as a unit for position tracking. Then, each entry is applied to the Large Table in sequence. If a crash occurs after the WAL write but before all Large Table updates complete, recovery replays the batch: the batch-start marker indicates how many entries follow, ensuring partial batches are reconstructed. This provides atomicity, either all or no update in the batch is visible after recovery.

\subsection{Read Flow} \label{sec:read}

\begin{figure}[t]
    \centering
    \includegraphics[width=0.8\columnwidth]{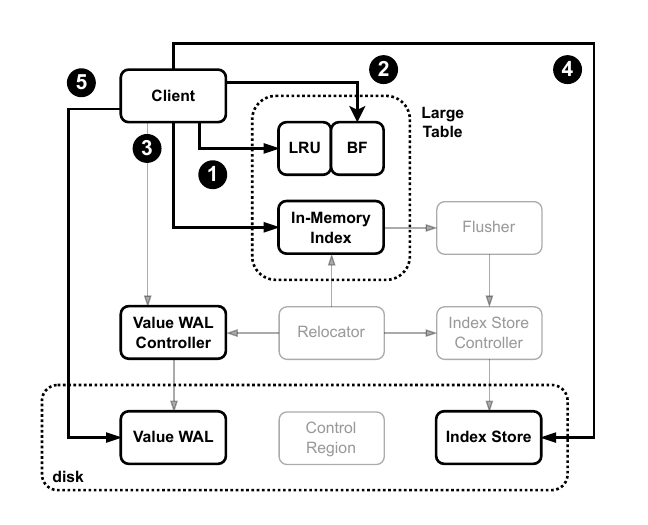}
    \Description{Read path diagram showing the tiered lookup strategy: first checking the LRU cache, then the Bloom filter, then the Large Table index (loading from Index Store if needed), and finally reading the value from the Value WAL.}
    \vskip -1em
    \caption{Read path showing the tiered lookup strategy. The snapshot component is omitted for clarity.}
    \label{fig:read}
    \vskip -1em
\end{figure}

\Cref{fig:read} shows the read path in \sysname by expanding the Large Table of \Cref{fig:architecture}. When a client issues a read operation, \sysname follows a tiered lookup strategy.

It first consults an in-memory LRU cache that stores recently accessed values~(\one). If the key is found, the value is returned immediately. This is the fastest path, requiring no index lookup or WAL access.
If the LRU cache misses, the system checks a bloom filter~(\two). If the bloom filter indicates the key is definitely not present, the lookup returns immediately, avoiding unnecessary index access. This is fundamentally different from RocksDB-style LSMs, where queries traverse multiple levels of SSTables regardless of whether the key exists.
If the bloom filter does not rule out the key, the system consults the Large Table~(\three). If the relevant index is resident in memory, the lookup completes immediately; otherwise, the system reads the index from the Index Store on disk~(\four). If the key is found, the index provides a WAL position, and the system reads the value directly from the Value WAL at that position~(\five). The value is then added to the LRU cache to accelerate future accesses. In practice, frequently accessed data remains in the operating system's page cache, making most reads memory-speed operations.

When a cell's index resides on disk rather than in memory, the system must typically load the entire index before performing a lookup. However, for uniformly distributed keys, \sysname can locate entries directly on disk without loading the full index. For large cells, this is wasteful when only a single key is needed. \sysname exploits a key property of its workload: keys are either cryptographic hashes or prefixed by such hashes, and thus uniformly distributed across the keyspace. This allows the system to estimate where a key would appear in a sorted, serialized index. The algorithm treats the key as an integer and computes its fractional position within the full key range, for example, a key starting with byte 0x80 lies roughly at the midpoint. Multiplying this fraction by the index file size yields an estimated byte offset.

The system then reads a small window around this offset. If the target key falls within the range covered by the window, binary search locates the entry. Otherwise, the window is shifted toward the beginning or end of the file based on key comparison, and the process repeats. For uniformly distributed keys, this typically converges within 1-3 iterations, reading only a few kilobytes rather than potentially megabytes of index data. We call this approach the \emph{optimistic index} and detail its statistical foundation in \Cref{sec:optimistic-index}.
This optimization is particularly valuable during recovery and cold-start scenarios, where most cells are unloaded. It allows \sysname to serve read requests immediately after opening the database, without waiting for indices to be loaded into memory (see \Cref{sec:recovery}). 

\subsection{State Snapshots} \label{sec:snapshot}

\sysname periodically captures snapshots in the background to enable efficient crash recovery. A snapshot records the state of the system at a point in time, allowing recovery to start from that point rather than replaying the entire Value WAL from the beginning. For each cell in the Large Table, a snapshot stores the Index Store position where the cell's most recently flushed index resides (if any) and a replay position indicating how much of the Value WAL is reflected in that flushed index. The snapshot also records a global replay-from position, the minimum replay position across all cells which marks the starting point for Value WAL replay during recovery.
The snapshot engine operates as a background thread that periodically flushes cells that have accumulated significant updates and writes this metadata atomically to the Control Region, a dedicated region on disk (\Cref{fig:architecture}). More frequent snapshots reduce recovery time by minimizing the Value WAL segment that must be replayed, but increase write I/O during normal operation; \sysname balances this trade-off by flushing only cells that exceed a configurable dirty threshold. Unlike LSM-trees based systems~\cite{rocksdb} that snapshot entire data structures, \sysname snapshots contain only positions rather than actual index data. As a result, they are small  and can be rewritten frequently without significant overhead.

\subsection{Crash Recovery} \label{sec:recovery}

When \sysname starts after a crash or normal shutdown, it performs a recovery sequence.
The system first reads the Control Region from disk. This contains the most recent snapshot: for each cell, the Index Store position where its flushed index resides, and a global replay-from position marking where Value WAL replay must begin.
The system then opens both the Value WAL and the Index Store. These append-only logs survived the crash; their contents remain intact.
The Large Table is initialized from the snapshot metadata. Each cell starts in an unloaded state, but the cell knows where its serialized index resides in the Index Store. This allows recovery to proceed without loading all indices upfront as they are loaded on demand during normal operation. Starting from the replay-from position recorded in the snapshot, the system iterates through the Value WAL and replays each entry into the Large Table, reconstructing any updates that occurred after the snapshot was taken.
Once replay completes, the system starts the background components: the Value WAL Controller, the Index Store Controller, the flusher threads, the relocator, and the snapshot thread. The database is now ready to serve client requests.
\section{Resource Management} \label{sec:management}

This section describes how \sysname manages resources to scale beyond available memory. We examine the Large Table's organization, which allows the index to partially reside on disk; index flushing, which persists updates to relieve memory pressure; and relocation, which reclaims disk space by pruning obsolete entries without blocking writes.

\subsection{The Large Table} \label{sec:large-table}

The Large Table mainly serves as \sysname's in-memory index, mapping keys to their corresponding positions in the Value WAL. Its design enables memory-efficient operation by allowing portions of the index to reside on disk while supporting concurrent access through fine-grained locking.

\paragraph{Partitioning and concurrency}
Supporting datasets larger than memory while maintaining high concurrency presents two challenges: the index cannot fit entirely in RAM, and a single lock would serialize all operations. \sysname addresses both by partitioning keys into cells, each representing a contiguous range of keys. This partitioning enables two key properties: cells can be individually loaded or evicted to manage memory, and cells can be locked independently to maximize concurrency.
Applications define one or more key spaces at database creation, each configured with a distribution type. For uniformly distributed keys (such as cryptographic hashes), cells are pre-allocated as a fixed-size array. For keys with common prefixes, cells are stored in a B-tree that grows dynamically. This latter organization suits composite keys such as (tenant\_id, record\_id) or (namespace, key), enabling efficient range scans within a prefix while accommodating new prefixes without pre-allocation.
Within each key space, cells are grouped into rows protected by sharded mutexes, allowing concurrent operations on different key ranges to proceed without contention.


\paragraph{Cell states}
A naive approach would load an entire cell's index into memory on first access, but this is prohibitive for write-heavy workloads: a single write to a cold cell could trigger loading megabytes of index data. \sysname avoids this with a \emph{DirtyUnloaded} state that buffers new writes in memory without loading the existing on-disk index.
Each cell maintains one of five states governing memory residency. A cell starts \emph{Empty} and transitions to \emph{Loaded} when it receives data that is fully resident in memory. When evicted under memory pressure, it becomes \emph{Unloaded}, with its index existing only on disk. \emph{DirtyLoaded} indicates an in-memory index with unpersisted changes, while \emph{DirtyUnloaded} buffers recent writes while the bulk of the index remains on disk.
When a write arrives for an Unloaded cell, the system transitions to DirtyUnloaded and buffers only the new entry. Reads check this in-memory buffer first; if the key is not found, the system performs a point lookup into the on-disk index without loading the entire structure (\Cref{sec:read}). This trades slightly higher per-lookup cost for substantially lower memory consumption.

\subsection{Optimistic Index Lookup} \label{sec:optimistic-index}

When a cell's index resides on disk, lookups typically require loading the index into memory or using a multi-level directory structure requiring multiple I/O operations. \sysname avoids both by exploiting a statistical property of uniform key distributions: the position of a key in a sorted array can be estimated from its value.

Consider $N$ keys sampled uniformly from $[0, 2^{256})$ and sorted into an array. For a key $k$, its expected position is $k \cdot N / 2^{256}$. More precisely, the $i$-th order statistic of $N$ uniform samples follows a Beta distribution, $\mathrm{Beta}(i, N-i+1)$, which provides confidence intervals for the key's location. This statistical foundation allows \sysname to bound the search region with high probability.

The on-disk index format is deliberately simple: a sorted array of fixed-size entries, each containing a 32-byte key and an 8-byte WAL position (40 bytes total). No headers, directories, or multi-level structures are needed. This simplicity enables direct position estimation: given a key $k$ and an index containing $N$ entries, the system computes the fractional position $p = k / 2^{256}$ and estimates the byte offset as $p \times 40N$.

To perform a lookup, the system reads a window of $W$ entries (800 by default, approximately 32\,KB) centered at the estimated offset. If the target key falls within the key range covered by the window, binary search locates the entry. If not, the window shifts toward the beginning or end of the file based on key comparison, and the process repeats. For uniformly distributed keys, this typically converges in one to three iterations.
The approach works because modern SSDs exhibit a batch read property: reading 32\,KB incurs essentially the same latency as reading a single byte, due to internal page sizes and command overhead. A window of 800 entries provides greater than 99\% probability of containing the target key for indices with 10K--100K entries. The window size is configurable, trading off between hit rate (larger windows miss less often) and I/O volume (smaller windows read fewer bytes per iteration). We evaluate this trade-off in \Cref{sec:eval}.

\subsection{Index Flushing} \label{sec:flushing}
When a cell accumulates too many dirty entries, the system must persist them to disk. A naive approach would block the cell during this flush, but this is unacceptable for write-heavy workloads where cells receive continuous updates. \sysname flushes asynchronously: the flusher captures a snapshot of the dirty index and dispatches persistence to background threads, while the cell remains available for new writes.

The flusher performs one of two operations depending on the cell's state: for DirtyLoaded cells, it serializes the complete in-memory index; for DirtyUnloaded cells, it first loads the existing on-disk index, merges it with the dirty entries, and serializes the combined result. The flusher may also remove obsolete entries.
Once the index is prepared, the flusher hands it to the Index Store Controller for persistence. The Index Store shares the same append-only implementation as the Value WAL (\Cref{sec:write}), differing only in the data it stores (serialized indices rather than key-value entries). It then notifies the Large Table with the new index position, so the cell can record where its on-disk index is stored for future reads. The cell performs an unmerge operation: it removes from its in-memory buffer all entries included in the flush, retaining only entries that arrived after the flush began. If no entries remain, the cell becomes to Unloaded; otherwise, it remains in DirtyUnloaded with the updated on-disk pointer.

A second challenge is serving reads during a flush. Because the Index Store is append-only, new indices are appended rather than overwriting existing ones. During a flush, the cell continues pointing to the previous index position, so concurrent reads safely access the old version. Only after the write completes does the flusher atomically update the cell's pointer. Readers and writers thus operate on disjoint regions, requiring no coordination beyond the final pointer update.

\subsection{Relocation}\label{sec:relocation}
As writes accumulate in the Value WAL, obsolete entries, overwritten values and deleted keys, consume space that can no longer be reclaimed by simply truncating the log. \sysname reclaims this space through a process called relocation, which moves live entries forward in the WAL, allowing old segments to be deleted.
Relocation operates by scanning the WAL, identifying entries that are still live, re-writing them to new positions at the tail of the WAL, updating the Large Table index to point to the new positions, and finally deleting the old WAL segments. The key challenge is ensuring correctness in the presence of concurrent writes: if a client updates a key while relocation is in progress, the relocated entry must not overwrite the newer value.
\sysname solves this with a compare-and-set mechanism. When relocation reads an entry at position $P$ and prepares to relocate it to position $P'$, it captures the last processed value $L$—the highest contiguous WAL position where all write operations have been written to the WAL and their corresponding index updates have been applied to the Large Table. When applying the update, it only replaces the index entry if the current position is less than $L$. If a concurrent write updated the key to position $P'' > L$, the relocation update is ignored, and the index continues to point to $P''$. This allows relocation to run concurrently with normal writes without locks or coordination.
Space is reclaimed at file granularity. The Value WAL is organized as a sequence of files; once all live entries in a file have been relocated (or were already beyond the file's range), the file can be deleted. The GC watermark tracks the oldest position that may still be referenced, ensuring files are not deleted prematurely.

\sysname supports two relocation strategies: \emph{WAL-based relocation} and \emph{index-based relocation}. WAL-based relocation scans the Value WAL sequentially from the oldest entry. For each entry, it checks whether the key still exists in the index and points to this WAL position—if so, the entry is live and must be relocated. This approach is simple and processes entries in WAL order, but must read each entry from disk.
Index-based relocation iterates through cells in the Large Table. For each cell, it loads the index, identifies entries whose WAL positions fall below the target cutoff (a configurable position that determines how much of the WAL to reclaim), reads their values from the WAL, and decides which to relocate. This approach can be more efficient when only a subset of cells contain old entries.
Both strategies support an optional relocation filter, an application-provided callback that decides for each entry whether to keep it, remove it, or stop relocation. This enables application-specific compaction logic, such as retaining only the latest version of each key or pruning entries based on business rules.

A key architectural difference between \sysname and LSM-tree databases like RocksDB is the role of compaction. In LSM-trees, compaction is essential for read performance: without it, reads must search through multiple levels of SSTables, and read amplification grows with the number of levels. Compaction merges levels, reducing the search space and keeping reads fast.
In \sysname, read performance is independent of WAL fragmentation. The Large Table index provides direct access to any value's WAL position—whether that position is at the beginning or end of the WAL, the read cost is the same (one index lookup plus one WAL read). Fragmentation does not degrade read latency.
This means relocation in \sysname serves a single purpose: reclaiming disk space. Unlike LSM-tree compaction, which repeatedly rewrites data across levels, \sysname writes each value to the WAL only once during normal operation. Additional writes occur only when relocation reclaims space, and even then, only live entries are rewritten. Because relocation is optional and can be deferred when disk space is abundant, applications can maintain near-$1\times$ write amplification.

Unlike LSM-tree compaction, which trades write amplification for read performance, \sysname's relocation trades write amplification purely for disk space. Applications with ample storage can defer or skip relocation entirely without impacting read latency.

\section{Implementation}\label{sec:implementation}

We provide a \emph{production-ready} implementation of \sysname in Rust~\cite{tidehunter-github}, currently being deployed in the Sui blockchain to replace RocksDB, comprising approximately 14,000 lines of code (excluding tests). The implementation leverages several key libraries that align with the system's design goals: \emph{memmap2} for memory-mapped I/O, enabling direct writes to WAL fragments without explicit system calls; \emph{parking\_lot} for high-performance mutexes with support for Arc-based guards, used in the guard-based position tracking mechanism; \emph{arc-swap} for atomic swapping of Arc pointers, allowing lock-free reads of shared data structures such as the WAL's memory map registry; \emph{crc32fast} for efficient CRC32 checksums on WAL entries; \emph{bloom} for per-cell Bloom filters that accelerate negative lookups; \emph{lru} for the value cache; and \emph{rayon} for parallel initialization of the Large Table during recovery. Unsafe code is limited to memory-mapped I/O operations and carefully encapsulated within the WAL module. The codebase includes approximately 160 unit tests, stress tests with shadow-state verification, and a failpoint infrastructure for deterministic concurrency testing.


\sysname employs a fixed set of background threads to handle asynchronous operations. Each WAL Controller (Value and Index) runs three dedicated threads: a \emph{tracker} that monitors position completion via the guard-based mechanism, a \emph{mapper} that manages memory-mapped region lifecycle and garbage collection, and a \emph{syncer} that issues fsync calls on finalized fragments. A configurable pool of \emph{flusher} threads handles index persistence, allowing multiple cells to be flushed concurrently. A single \emph{relocator} thread performs background compaction, moving live entries forward in the Value WAL to reclaim space. Finally, a \emph{snapshot} thread periodically captures consistent checkpoints of the Large Table for crash recovery. In total, a typical deployment runs 8 background threads plus the configured number of flushers. The design deliberately avoids per-client or per-request thread spawning to minimize scheduling overhead and maintain predictable resource usage.

\section{Experimental Evaluation}\label{sec:eval}

We answer the following questions about \sysname:
\begin{enumerate}
    \item How does the read-write ratio influence the performance of \sysname?
    \item How does the type of read operation (get, exists, scan) influence the performance of \sysname?
    \item How does workload skew influence the performance of \sysname?
    \item How does value size influence the performance of \sysname?
    \item How does relocation influence the performance of \sysname?
    \item How does our optimistic index perform against the naive index?
\end{enumerate}

\subsection{Experimental Setup}

We run our experiments on a Medium v4 OpenMetal instance~\cite{openmetal} equipped with two Intel Xeon Silver 4510 processors providing 24 CPU cores (48 threads) at 2.4--4.1\,GHz, 256\,GB of DDR5 4400\,MHz RAM, and a 6.4\,TB NVMe drive.

\paragraph{Baselines.}
We compare \sysname against the following state-of-the-art key-value stores:
\begin{itemize}
    \item \textbf{RocksDB}~\cite{rocksdb}: An LSM-tree-based embedded key-value store widely used in production systems.
    \item \textbf{BlobDB}~\cite{blobdb}: RocksDB's integrated blob storage extension inspired by WiscKey~\cite{lu2016wisckey}, which separates large values from the LSM-tree to reduce write amplification.
\end{itemize}

\paragraph{Methodology.}
We evaluate \sysname against the baselines using a custom benchmark consisting of two phases: a \emph{fill phase} and a \emph{measurement phase}.
In the fill phase, we populate the database with random keys and values.
We use a fixed key size of 32 bytes, while the value size is a configurable parameter.
We insert entries until the database reaches 1\,TB of data; specifically, we divide 1\,TB by the entry size (key size plus value size) to determine the number of insertions.
After the fill phase, we allow a 10-minute cooldown period for the system to stabilize (e.g., for relocation to settle).
In the measurement phase, we perform operations for a fixed duration of 10 minutes.
The following workload parameters are configurable:
\begin{itemize}
    \item \textbf{Read-write ratio}: We evaluate three configurations---100\% writes, 50\% reads with 50\% writes, and 100\% reads.
    \item \textbf{Read operation type}: We consider three read operations: \emph{get} (returns the value for a key if it exists), \emph{exists} (returns whether a key exists), and \emph{reverse iterator} (returns the key-value pair with the largest key smaller than the query key, if one exists).
    \item \textbf{Skew}: A homogeneous workload selects keys uniformly at random, while a skewed workload favors more recently inserted keys. We implement skew using a Zipfian distribution with parameter $\theta = 0$ (homogeneous) or $\theta = 2$ (skewed). We use ``homogeneous'' rather than ``uniform'' for access patterns to avoid confusion with uniformly distributed keys.
    \item \textbf{Value size}: We evaluate both large values (1\,KB) and small values (64 bytes and 128 bytes).
    \item \textbf{Relocation}: Relocation can be enabled or disabled to measure its impact on performance. Unless stated otherwise, all benchmark results were obtained with relocation enabled.
\end{itemize}

\subsection{Benchmark Results}

We present benchmark results for \sysname, RocksDB, and BlobDB across various workload configurations.

\subsubsection{Large Values}
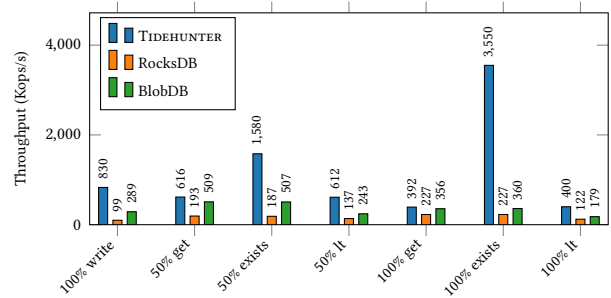
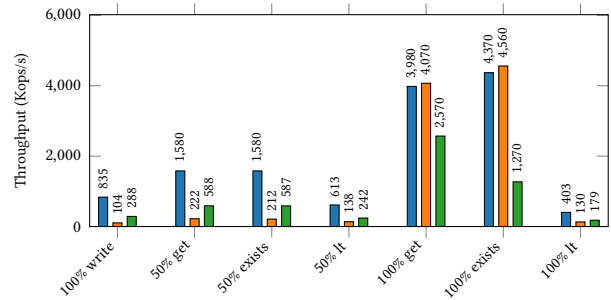
\begin{figure}[t]
    \centering
    \subfloat[Homogeneous Distribution ($\theta=0$)]{%
        \begin{tikzpicture}
            \begin{axis}[widebarplot, ymax=4000]
            \addplot[fill=tidehunterblue] coordinates
                {(100\% write,830) (50\% get,616) (50\% exists,1580) (50\% lt,612) (100\% get,392) (100\% exists,3550) (100\% lt,400)};
            \addplot[fill=rocksdborange] coordinates
                {(100\% write,99) (50\% get,193) (50\% exists,187) (50\% lt,137) (100\% get,227) (100\% exists,227) (100\% lt,122)};
            \addplot[fill=blobdbgreen] coordinates
                {(100\% write,289) (50\% get,509) (50\% exists,507) (50\% lt,243) (100\% get,356) (100\% exists,360) (100\% lt,179)};
            \legend{\sysname, RocksDB, BlobDB}
            \end{axis}
        \end{tikzpicture}%
        \label{fig:homogeneous}%
    }\\[2ex]
    \subfloat[Skewed Distribution ($\theta=2$)]{%
        \begin{tikzpicture}
            \begin{axis}[widebarplot, ymax=5100]
            \addplot[fill=tidehunterblue] coordinates
                {(100\% write,835) (50\% get,1580) (50\% exists,1580) (50\% lt,613) (100\% get,3980) (100\% exists,4370) (100\% lt,403)};
            \addplot[fill=rocksdborange] coordinates
                {(100\% write,104) (50\% get,222) (50\% exists,212) (50\% lt,138) (100\% get,4070) (100\% exists,4560) (100\% lt,130)};
            \addplot[fill=blobdbgreen] coordinates
                {(100\% write,288) (50\% get,588) (50\% exists,587) (50\% lt,242) (100\% get,2570) (100\% exists,1270) (100\% lt,179)};
            \end{axis}
        \end{tikzpicture}%
        \label{fig:skewed}%
    }
    \Description{Two bar charts comparing throughput of Tidehunter, RocksDB, and BlobDB for 1KB values. Top chart shows homogeneous distribution where Tidehunter dominates with 830K ops/s for writes. Bottom chart shows skewed distribution where performance converges for read-heavy workloads.}
    \caption{\footnotesize Throughput comparison across workloads. The 50\% read cases include 50\% writes.
             Top: homogeneous access distribution ($\theta=0$). Bottom: skewed distribution ($\theta=2$).}
    \label{fig:benchmark-results-1k}
\end{figure}

The results for large values are in \Cref{fig:benchmark-results-1k}.

\paragraph{Write Performance.}
Under write-only workloads (\Cref{fig:homogeneous}, leftmost bars), \sysname achieves 830K ops/sec, outperforming RocksDB by 8.4$\times$ and BlobDB by 2.9$\times$. This substantial advantage stems from \sysname's append-only WAL design: each write requires only a single sequential disk operation, whereas LSM-tree-based systems must eventually compact and rewrite data multiple times. BlobDB partially mitigates this overhead by separating large values from the LSM-tree, explaining its intermediate performance.

\paragraph{Read Performance.}
For homogeneous read workloads, \sysname maintains a consistent advantage across all operation types. On get operations (\Cref{fig:homogeneous}), \sysname achieves 392K ops/s compared to RocksDB's 227K ops/s, a 1.7$\times$ improvement. The exists operation (\Cref{fig:homogeneous}) reveals an even more pronounced advantage: \sysname reaches 3.55M ops/s, outperforming RocksDB by 15.6$\times$. This dramatic difference occurs because \sysname can resolve existence queries directly from the index without fetching the actual value, while LSM-trees must traverse multiple levels regardless of whether the value is needed.

For reverse iterator operations (\Cref{fig:homogeneous}), \sysname maintains a 3.3$\times$ advantage over RocksDB (400K vs 122K ops/s). However, all systems show reduced throughput compared to point queries, as range operations require examining multiple entries.

\paragraph{Mixed Workloads.}
At a 50/50 read-write ratio, \sysname continues to lead with 616K ops/s for get operations, compared to RocksDB's 193K ops/s. The performance gap narrows relative to write-only workloads but remains substantial (3.2$\times$), demonstrating that \sysname's design also benefits mixed read-write paths.

\paragraph{Effect of Skew.}
Under skewed workloads (\Cref{fig:skewed}), where accesses favor recently inserted keys, all systems benefit from caching effects. For read-only get operations with skew, RocksDB achieves 4.07M ops/s, approaching \sysname's 3.98M ops/s. This convergence occurs because hot data resides in memory across all systems---RocksDB's block cache and \sysname's large table both serve frequently accessed entries without disk I/O. However, \sysname maintains its advantage for write-heavy and mixed workloads even under skew, as its write path remains fundamentally more efficient.

\paragraph{Summary.}
For large values, \sysname consistently outperforms LSM-tree-based alternatives. The advantage is most pronounced for write-heavy workloads (up to 8.4$\times$) and existence checks (up to 15.6$\times$), and remains significant for read-heavy homogeneous workloads (1.7$\times$). These results validate \sysname's design goal of minimizing data movement.
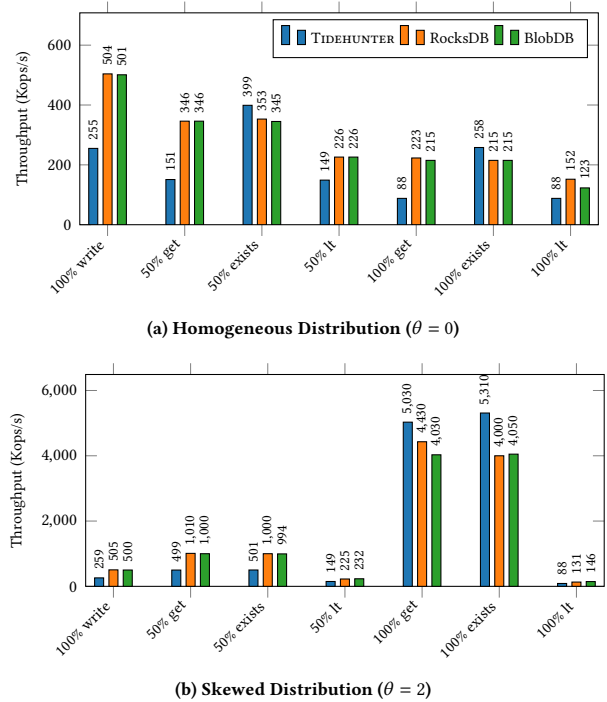
\begin{figure}[t]
    \centering
    \subfloat[Homogeneous Distribution ($\theta=0$)]{%
        \begin{tikzpicture}
            \begin{axis}[widebarplot, ymax=600]
            \addplot[fill=tidehunterblue] coordinates
                {(100\% write,255) (50\% get,151) (50\% exists,399) (50\% lt,149) (100\% get,88) (100\% exists,258) (100\% lt,88)};
            \addplot[fill=rocksdborange] coordinates
                {(100\% write,504) (50\% get,346) (50\% exists,353) (50\% lt,226) (100\% get,223) (100\% exists,215) (100\% lt,152)};
            \addplot[fill=blobdbgreen] coordinates
                {(100\% write,501) (50\% get,346) (50\% exists,345) (50\% lt,226) (100\% get,215) (100\% exists,215) (100\% lt,123)};
            \legend{\sysname, RocksDB, BlobDB}
            \pgfplotsset{legend columns=3, legend style={at={(0.98,0.97)}, anchor=north east}}
            \end{axis}
        \end{tikzpicture}%
        \label{fig:homogeneous-64b}%
    }\\[2ex]
    \subfloat[Skewed Distribution ($\theta=2$)]{%
        \begin{tikzpicture}
            \begin{axis}[widebarplot, ymax=5500]
            \addplot[fill=tidehunterblue] coordinates
                {(100\% write,259) (50\% get,499) (50\% exists,501) (50\% lt,149) (100\% get,5030) (100\% exists,5310) (100\% lt,88)};
            \addplot[fill=rocksdborange] coordinates
                {(100\% write,505) (50\% get,1010) (50\% exists,1000) (50\% lt,225) (100\% get,4430) (100\% exists,4000) (100\% lt,131)};
            \addplot[fill=blobdbgreen] coordinates
                {(100\% write,500) (50\% get,1000) (50\% exists,994) (50\% lt,232) (100\% get,4030) (100\% exists,4050) (100\% lt,146)};
            \end{axis}
        \end{tikzpicture}%
        \label{fig:skewed-64b}%
    }
    \Description{Two bar charts comparing throughput of Tidehunter, RocksDB, and BlobDB for 64-byte values. Top chart shows homogeneous distribution where RocksDB leads. Bottom chart shows skewed distribution where Tidehunter catches up for read-heavy workloads due to caching.}
    \caption{\footnotesize Throughput comparison for small values (64 bytes). The 50\% read cases include 50\% writes.
             Top: homogeneous access distribution ($\theta=0$). Bottom: skewed distribution ($\theta=2$).}
    \label{fig:benchmark-results-64b}
\end{figure}

\subsubsection{Small Values}
The results for small values are in \Cref{fig:benchmark-results-64b}.
For small values, the performance characteristics differ substantially from the large-value case, revealing the trade-offs inherent in \sysname's design.

\paragraph{Write Performance.}
Under write-only workloads (\Cref{fig:homogeneous-64b}, leftmost bars), RocksDB and BlobDB achieve approximately 500K ops/s, outperforming \sysname's 255K ops/s by nearly 2$\times$. This reversal occurs because the overhead of \sysname's separate index becomes proportionally more significant when values are small. LSM-trees amortize their metadata costs across sorted runs and benefit from efficient compression of small, collocated entries.

\paragraph{Read Performance.}
The gap widens for read operations under homogeneous access patterns. For get operations at 100\% reads (\Cref{fig:homogeneous-64b}, rightmost bars), RocksDB achieves 223K ops/s compared to \sysname's 88K ops/s---a 2.5$\times$ advantage. This difference reflects LSM-trees' superior cache efficiency: sorted runs pack many small entries into each cached block, whereas \sysname's position-based lookups may incur more cache misses when entries are scattered across the WAL.
Notably, \sysname's exists operation (\Cref{fig:homogeneous-64b}) shows improved relative performance, achieving 258K ops/s at 100\% reads compared to RocksDB's 215K ops/s. This demonstrates that \sysname's existence check optimization provides value even for small entries.

\paragraph{Effect of Skew.}
Skewed workloads (\Cref{fig:skewed-64b}) reveal an important exception to LSM-trees' small-value advantage. Under read-only workloads with high skew, \sysname achieves 5.03M ops/s for get operations, surpassing RocksDB's 4.43M ops/s by 14\%. For exists operations, \sysname's advantage grows to 33\% (5.31M vs 4.00M ops/s). This reversal occurs because \sysname's in-memory large table efficiently caches hot entries, and the WAL's append-only structure means recently written (and frequently accessed under skew) entries are contiguous, improving cache locality.

\paragraph{Summary.}
For small 64-byte values with homogeneous access patterns, LSM-tree-based systems outperform \sysname by 2--2.5$\times$. However, \sysname regains the advantage under skewed read workloads where caching effects dominate. This suggests that \sysname is better suited for workloads with larger values or significant temporal locality, while LSM-trees remain preferable for small-value workloads with homogeneous access patterns.

\vskip 0.5em

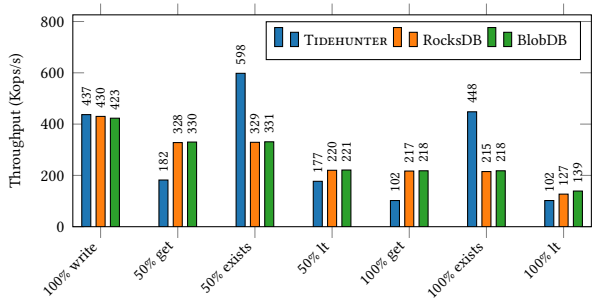
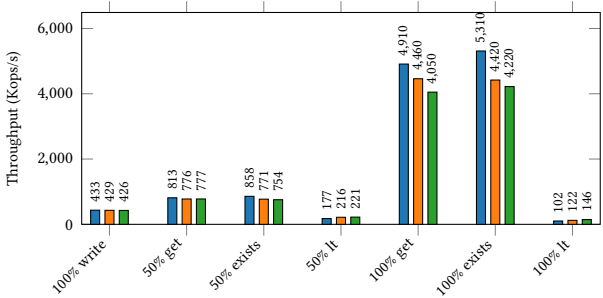
\begin{figure}[t]
    \centering
    \subfloat[Homogeneous Distribution ($\theta=0$)]{%
        \begin{tikzpicture}
            \begin{axis}[widebarplot, ymax=700]
            \addplot[fill=tidehunterblue] coordinates
                {(100\% write,437) (50\% get,182) (50\% exists,598) (50\% lt,177) (100\% get,102) (100\% exists,448) (100\% lt,102)};
            \addplot[fill=rocksdborange] coordinates
                {(100\% write,430) (50\% get,328) (50\% exists,329) (50\% lt,220) (100\% get,217) (100\% exists,215) (100\% lt,127)};
            \addplot[fill=blobdbgreen] coordinates
                {(100\% write,423) (50\% get,330) (50\% exists,331) (50\% lt,221) (100\% get,218) (100\% exists,218) (100\% lt,139)};
            \legend{\sysname, RocksDB, BlobDB}
            \pgfplotsset{legend columns=3, legend style={at={(0.98,0.97)}, anchor=north east}}
            \end{axis}
        \end{tikzpicture}%
        \label{fig:homogeneous-128b}%
    }\\[2ex]
    \subfloat[Skewed Distribution ($\theta=2$)]{%
        \begin{tikzpicture}
            \begin{axis}[widebarplot, ymax=5500]
            \addplot[fill=tidehunterblue] coordinates
                {(100\% write,433) (50\% get,813) (50\% exists,858) (50\% lt,177) (100\% get,4910) (100\% exists,5310) (100\% lt,102)};
            \addplot[fill=rocksdborange] coordinates
                {(100\% write,429) (50\% get,776) (50\% exists,771) (50\% lt,216) (100\% get,4460) (100\% exists,4420) (100\% lt,122)};
            \addplot[fill=blobdbgreen] coordinates
                {(100\% write,426) (50\% get,777) (50\% exists,754) (50\% lt,221) (100\% get,4050) (100\% exists,4220) (100\% lt,146)};
            \end{axis}
        \end{tikzpicture}%
        \label{fig:skewed-128b}%
    }
    \Description{Two bar charts comparing throughput of Tidehunter, RocksDB, and BlobDB for 128-byte values. This represents the crossover point where performance is comparable for write-heavy workloads, with Tidehunter excelling at existence checks.}
    \caption{\footnotesize Throughput comparison for small values (128 bytes). The 50\% read cases include 50\% writes.
             Top: homogeneous access distribution ($\theta=0$). Bottom: skewed distribution ($\theta=2$).}
    \label{fig:benchmark-results-128b}
\end{figure}
The results for 128-byte values are in \Cref{fig:benchmark-results-128b}.
The 128-byte value size represents a transitional point between small and large value performance characteristics, where neither architecture holds a decisive advantage across all workloads.

\paragraph{Write Performance.}
Under write-only workloads (\Cref{fig:homogeneous-128b}, leftmost bars), all systems achieve comparable throughput: \sysname at 437K ops/s, RocksDB at 430K ops/s, and BlobDB at 423K ops/s. This near-parity suggests that 128 bytes approaches the crossover point where \sysname's single-write advantage begins to offset LSM-trees' efficient small-entry handling.

\paragraph{Read Performance.}
For homogeneous read workloads, LSM-trees maintain an advantage for value-fetching operations. At 100\% reads, RocksDB achieves 217K ops/s for get operations compared to \sysname's 102K ops/s (\Cref{fig:homogeneous-128b}). However, \sysname excels at existence checks: at a 50/50 read-write ratio, \sysname achieves 598K ops/s for exists operations, outperforming RocksDB's 329K ops/s by 1.8$\times$ (\Cref{fig:homogeneous-128b}). This demonstrates that \sysname's optimized existence check provides substantial benefits.

For reverse iterator operations (\Cref{fig:homogeneous-128b}), \sysname leads under write-heavy workloads (439K vs 382K ops/s) but trails under read-heavy workloads (102K vs 127K ops/s), reflecting the mixed nature of this transitional value size.

\paragraph{Effect of Skew.}
Under skewed access patterns (\Cref{fig:skewed-128b}), \sysname's in-memory caching provides a consistent advantage for read-heavy workloads. For get operations at 100\% reads with skew, \sysname achieves 4.91M ops/s versus RocksDB's 4.46M ops/s (10\% improvement). The advantage is more pronounced for exists operations, where \sysname reaches 5.31M ops/s compared to RocksDB's 4.42M ops/s (20\% improvement).

\paragraph{Summary.}
At 128 bytes, \sysname and LSM-tree systems show more balanced performance compared to smaller values. \sysname matches or exceeds LSM-trees for write-heavy workloads and existence checks, while LSM-trees retain an advantage for read-heavy homogeneous access patterns. This transitional behavior indicates that \sysname's advantages grow progressively as value sizes increase beyond this threshold.

\subsubsection{Effect of Relocation}
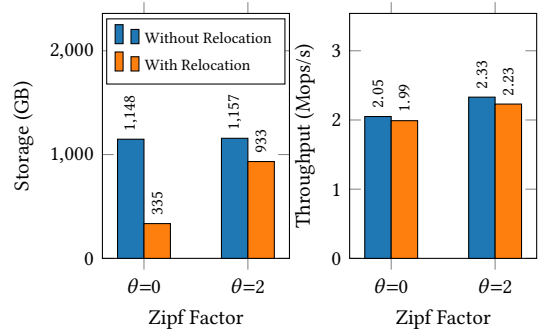
\begin{figure}[t]
    \centering
    \begin{tikzpicture}
        \begin{groupplot}[
            group style={
                group size=2 by 1,
                horizontal sep=0.8cm,
            },
            ybar=0pt,
            enlarge x limits=0.4,
            symbolic x coords={$\theta{=}0$,$\theta{=}2$},
            xtick=data,
            ymin=0,
            enlarge y limits={upper,value=0.18},
            legend style={at={(0.02,0.98)}, anchor=north west, font=\scriptsize, cells={anchor=west}},
            nodes near coords,
            every node near coord/.append style={font=\scriptsize, rotate=90, anchor=west},
            width=0.48\columnwidth,
            height=0.22\textheight,
            tick label style={font=\small},
            label style={font=\small},
        ]
        \nextgroupplot[ylabel={Storage (GB)}, xlabel={Zipf Factor}, ymax=2000]
        \addplot[fill=tidehunterblue] coordinates
            {($\theta{=}0$,1148) ($\theta{=}2$,1157)};
        \addplot[fill=rocksdborange] coordinates
            {($\theta{=}0$,335) ($\theta{=}2$,933)};
        \legend{Without Relocation, With Relocation}
        
        \nextgroupplot[ylabel={Throughput (Mops/s)}, xlabel={Zipf Factor}, ymax=3.0]
        \addplot[fill=tidehunterblue] coordinates
            {($\theta{=}0$,2.05) ($\theta{=}2$,2.33)};
        \addplot[fill=rocksdborange] coordinates
            {($\theta{=}0$,1.99) ($\theta{=}2$,2.23)};
        \end{groupplot}
    \end{tikzpicture}
    \Description{Two side-by-side bar charts comparing storage and throughput with and without relocation. Left chart shows storage reduced from 1148GB to 335GB for uniform distribution and from 1157GB to 933GB for skewed distribution. Right chart shows throughput is similar with and without relocation.}
    \caption{\footnotesize Effect of relocation on storage usage and throughput. Left: storage after mixed phase. Right: throughput during mixed phase. Relocation significantly reduces storage with minimal throughput impact.}
    \label{fig:relocation-results}
\end{figure}
To evaluate relocation's impact on storage efficiency and performance, we ran a delete-heavy workload (100\% deletes) under both uniform ($\theta{=}0$) and skewed ($\theta{=}2$) access patterns (\Cref{fig:relocation-results}), after the database has been pre-filled with 1\,TB of data. We used large values (1\,KB).
Without relocation, deleted entries leave gaps that accumulate, consuming 1148\,GB and 1157\,GB respectively.
With relocation enabled, storage drops to 335\,GB (uniform) and 933\,GB (skewed)---a 71\% and 19\% reduction.
The smaller savings under skew occur because only frequently accessed keys are deleted, leaving fewer reclaimable gaps.
Throughput overhead remains minimal: 3\% for uniform and 4\% for skewed workloads.

\subsection{Index Format Comparison}

We microbenchmark the lookup performance of \sysname's two persistent index formats: the optimistic index and a naive baseline.
%
\begin{itemize}
    \item \textbf{Header-based index}: This format uses a fixed-size header of 128 entries (1\,KB total) to partition keys into micro-cells based on their prefix. Each header entry stores offsets pointing to a contiguous region of sorted key-value pairs. A lookup requires exactly two I/O operations: reading the 8-byte header entry to determine the data region, followed by reading and binary searching the data region.
    \item \textbf{Optimistic index} (\Cref{sec:optimistic-index}): This format assumes a uniform key distribution to estimate the probable location of a key within the sorted index file. It uses a window-based search strategy with a default window size of 800 entries. The search begins at the estimated offset and iteratively moves the window until the target key falls within bounds, then performs a binary search. This approach eliminates header overhead and can achieve single-roundtrip lookups when the distribution estimate is accurate.
\end{itemize}

\paragraph{Data Generation.}
We generate 25\,000 separate index files, each containing 1\,000\,000 randomly populated entries. Each entry consists of a 32-byte key generated uniformly at random and an 8-byte WAL position. The indices are serialized to disk using their respective formats, resulting in approximately 1\,TB of  data per format.

\paragraph{Benchmark Configuration.}
The benchmark performs 1\,000\,000 lookups per run across 10 independent runs. Lookups are grouped into batches of 1\,000 operations for timing measurements. The window size for the optimistic index is set to 800 entries. We evaluate both single- and multi-threaded configurations, and optionally enable direct I/O to bypass the operating system's page cache.

\paragraph{Workload.}
The benchmark uses a random access pattern where each lookup selects a uniformly random index from the 25\,000 available indices and queries a randomly generated 32-byte key. Since the lookup keys are generated independently of the keys stored in the indices, this workload consists predominantly of negative lookups (keys not present in the index), which represents a worst-case scenario for the optimistic index as it may require multiple window movements to determine key absence.

\begin{figure*}[t]
    \centering
    \ref{indexwindowlegend}\\[1ex]
    \begin{tikzpicture}
        \begin{groupplot}[
            group style={
                group size=3 by 1,
                horizontal sep=0.5cm,
                ylabels at=edge left,
            },
            indexwindowplot,
            width=0.38\textwidth,
            xmode=log,
            log basis x=2,
        ]
        \nextgroupplot[title={(a) 1 thread}, ymax=12]
            \addplot[color=headerblue, mark=o, solid] coordinates {(100,5.89) (200,5.88) (400,5.88) (800,5.88) (1600,5.88) (3200,5.88)};
            \addplot[color=headerblue, mark=o, dashed] coordinates {(100,9.69) (200,9.46) (400,9.33) (800,9.35) (1600,9.32) (3200,9.27)};
            \addplot[color=optimisticorange, mark=square, solid] coordinates {(100,3.08) (200,4.81) (400,6.82) (800,8.53) (1600,9.54) (3200,6.90)};
            \addplot[color=optimisticorange, mark=square, dashed] coordinates {(100,4.51) (200,5.99) (400,7.15) (800,7.92) (1600,8.04) (3200,6.50)};

        \nextgroupplot[title={(b) 16 threads}, ymax=120, legend to name=indexwindowlegend, legend columns=4]
            \addplot[color=headerblue, mark=o, solid] coordinates {(100,76.60) (200,76.63) (400,76.47) (800,76.51) (1600,76.58) (3200,76.61)};
            \addplot[color=headerblue, mark=o, dashed] coordinates {(100,67.30) (200,66.65) (400,63.22) (800,65.31) (1600,68.37) (3200,68.38)};
            \addplot[color=optimisticorange, mark=square, solid] coordinates {(100,44.87) (200,67.72) (400,89.26) (800,99.98) (1600,91.07) (3200,52.80)};
            \addplot[color=optimisticorange, mark=square, dashed] coordinates {(100,62.21) (200,75.77) (400,75.23) (800,67.50) (1600,51.12) (3200,30.31)};
            \legend{Header (DIO), Header (no DIO), Optimistic (DIO), Optimistic (no DIO)}

        \nextgroupplot[title={(c) 48 threads}, ymax=220]
            \addplot[color=headerblue, mark=o, solid] coordinates {(100,156.34) (200,156.38) (400,156.37) (800,156.39) (1600,156.33) (3200,156.40)};
            \addplot[color=headerblue, mark=o, dashed] coordinates {(100,56.47) (200,56.63) (400,56.85) (800,56.68) (1600,56.51) (3200,56.41)};
            \addplot[color=optimisticorange, mark=square, solid] coordinates {(100,114.98) (200,163.61) (400,193.82) (800,161.70) (1600,102.84) (3200,54.95)};
            \addplot[color=optimisticorange, mark=square, dashed] coordinates {(100,68.99) (200,70.90) (400,65.50) (800,54.70) (1600,38.52) (3200,23.05)};
        \end{groupplot}
    \end{tikzpicture}
    \Description{Three line charts showing throughput versus window size (100 to 3200) for 1, 16, and 48 threads. Header-based index shows flat lines since window size does not affect it. Optimistic index shows a peak at intermediate window sizes, with optimal size decreasing as thread count increases.}
    \caption{\footnotesize Index lookup throughput vs.\ window size for different thread counts. The header-based index maintains constant throughput (flat lines), while the optimistic index performance varies with window size. The optimal window size decreases as concurrency increases.}
    \label{fig:index-throughput-window}
\end{figure*}
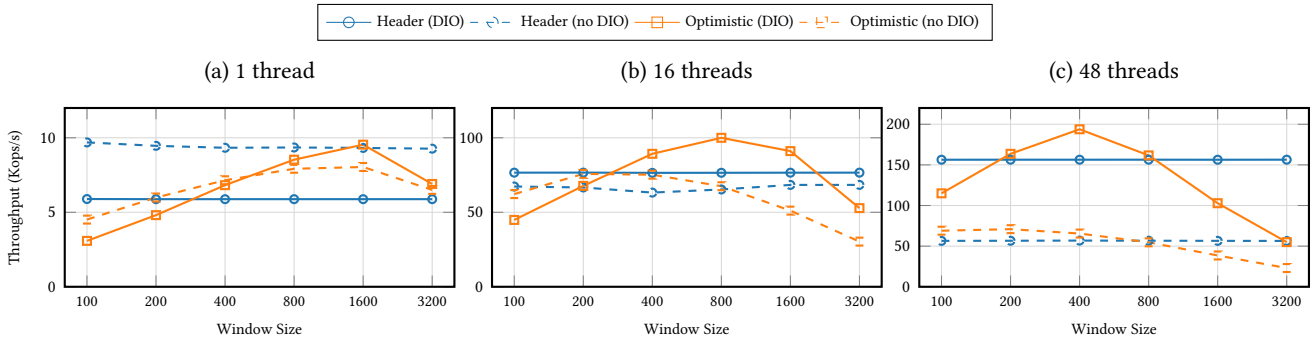

\paragraph{Results.}
\Cref{fig:index-throughput-window} shows the throughput scaling behavior of both index formats as a function of thread count. For the optimistic index, we use the optimal window size for each configuration (which varies from 1600 at low thread counts to 200--400 at high thread counts). For the header-based index, window size does not affect its performance.

With direct I/O enabled, the optimistic index outperforms the header-based index at all thread counts, achieving up to 194K lookups/s at 48 threads compared to 156K for the header-based index---a 24\% improvement. Without direct I/O, both indices plateau at high thread counts due to page cache contention, with the optimistic index achieving 71K lookups/s compared to 57K for the header-based index.

\Cref{fig:index-throughput-window} shows the sensitivity of the optimistic index to window size across different thread counts. The header-based index maintains constant throughput regardless of window size (since it always performs exactly two I/O operations per lookup), while the optimistic index exhibits a clear trade-off: smaller windows require more iterations but read less data per iteration, while larger windows reduce iterations but increase I/O volume. The optimal window size decreases as thread count increases, reflecting the benefits of smaller I/O operations under high concurrency.

\subsection{Case Study: Sui Blockchain}

To demonstrate real-world performance, we integrated \sysname into the Sui blockchain, which currently uses RocksDB as its database backend.
We replaced Sui's validator storage layer with \sysname to evaluate performance impact and demonstrate production readiness.
We deployed Sui in a test cluster of 150 geographically distributed machines under an artificial load of 6,000 transactions per second.

\begin{figure}[t]
    \centering
    \includegraphics[width=\columnwidth]{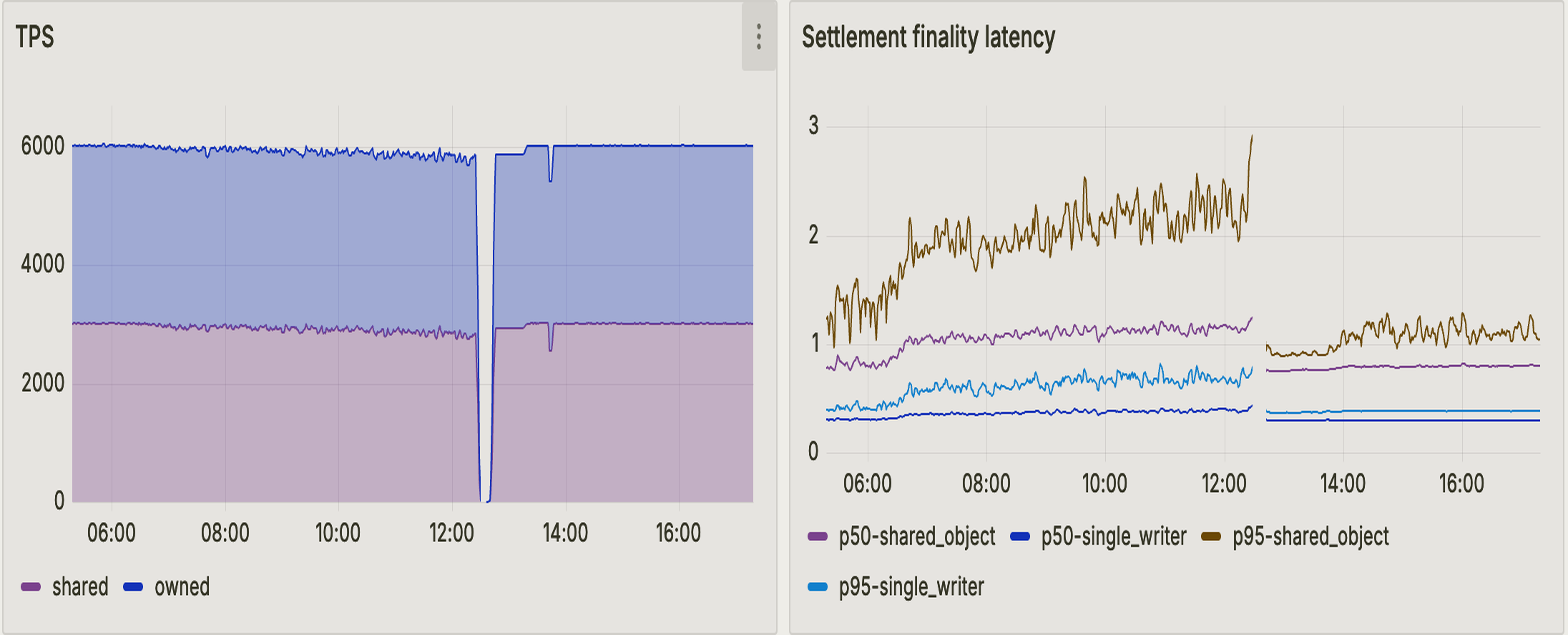}
    \Description{Side-by-side comparison of transaction latency over time for RocksDB (before 12:30) and TideHunter (after 12:30) on Sui blockchain. RocksDB shows increasing latency after several hours while TideHunter maintains stable latency.}
    \caption{\footnotesize Transaction latency comparison on Sui blockchain under 6,000 TPS load. RocksDB (before 12:30) shows latency degradation over time.  \sysname (after 12:30) maintains stable latency.}
    \label{fig:sui-latency}
\end{figure}

\begin{figure}[t]
    \centering
    \includegraphics[width=\columnwidth]{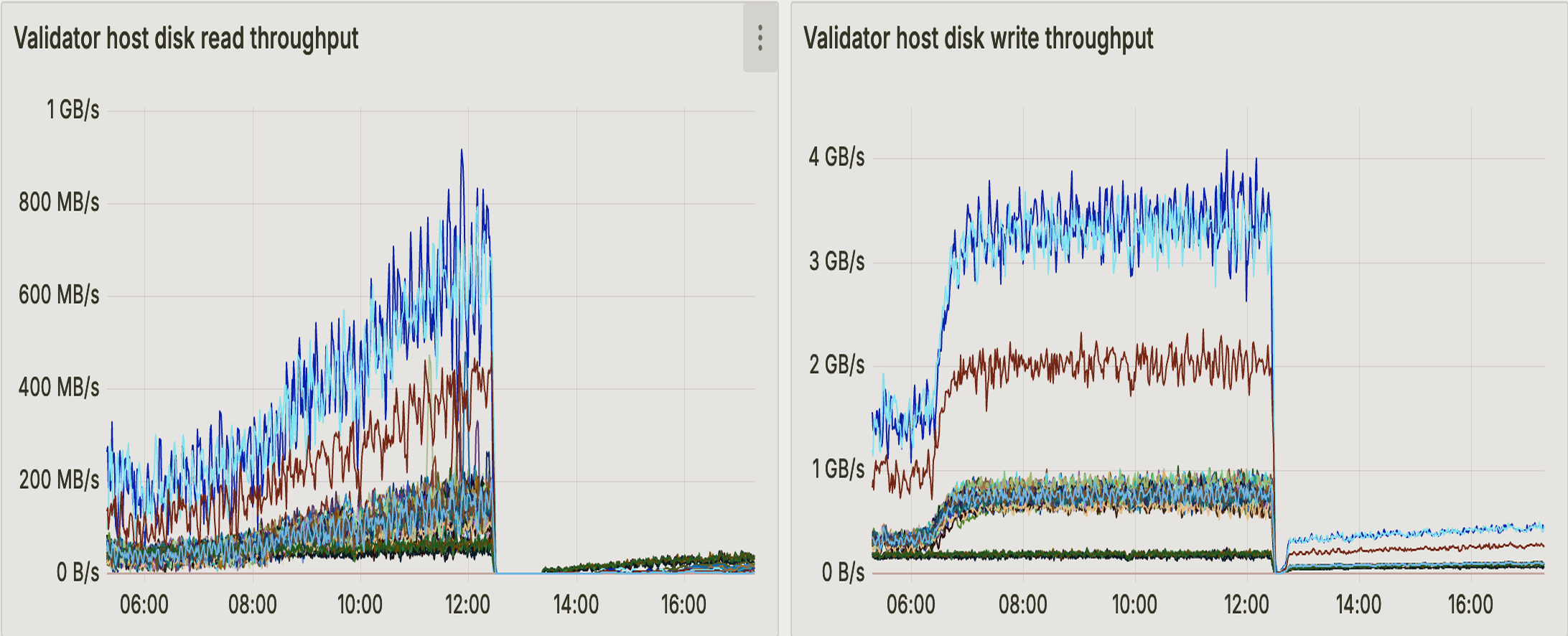}
    \Description{Side-by-side comparison of disk read and write rates for RocksDB (before 12:30) and TideHunter (after 12:30) on Sui blockchain. TideHunter shows significantly lower disk utilization.}
    \caption{\footnotesize Disk I/O utilization comparison on Sui blockchain under 6000 TPS load. \sysname (after 12:30) achieves significantly lower disk read and write rates than RocksDB (before 12:30).}
    \label{fig:sui-disk}
\end{figure}

All hosts started the experiment using RocksDB and at 12:30 switched to using \sysname. These results are presented in \Cref{fig:sui-latency,fig:sui-disk}.
The first thing to notice is that under the sustained load of 6,000 TPS, RocksDB performance degrades after several hours, resulting in significant latency increases (\Cref{fig:sui-latency} before 12:30). In contrast, once \sysname is deployed the same hosts maintain stable latency throughout the test period (\Cref{fig:sui-latency}, after 12:30).
The same characteristics are also present in the  disk I/O utilization (\Cref{fig:sui-disk}), where the high write amplification of RocksDB results in 1-4GB/s write throughput demand, whereas the need to search deep inside the LSM-Tree results in 0.2-0.8 GB/s read throughput demand. In contrast, once \sysname is enabled the demand of the disk resource drops to less than 100MB/s for both reads and writes.

\begin{figure}[t]
    \centering
    \includegraphics[width=\columnwidth]{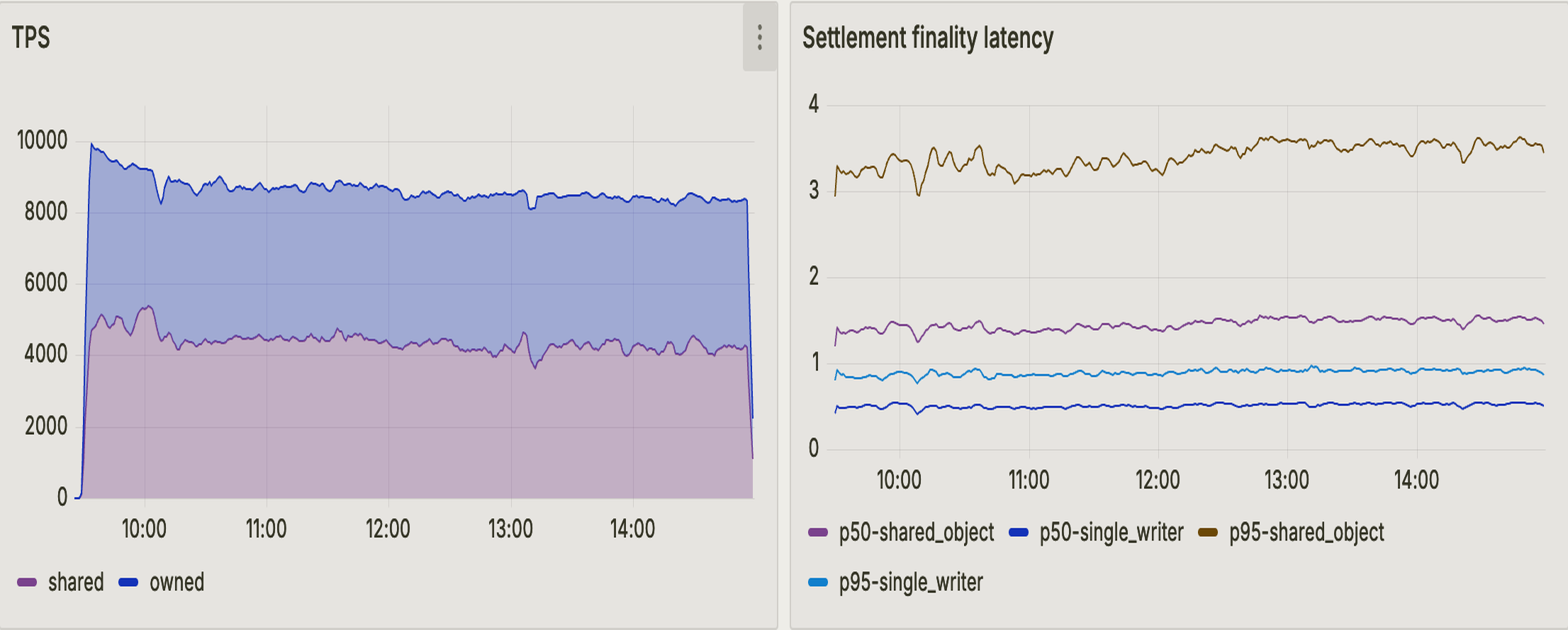}
    \Description{Performance of Sui using \sysname under 8500 TPS load. Metrics remain steady, unlike RocksDB with 6000 TPS.}
    \caption{\footnotesize Performance of Sui using \sysname under 8500 TPS load. Metrics remain steady, unlike RocksDB with 6000 TPS (\Cref{fig:sui-latency}, before 12:30).}
    \label{fig:sui-throughput}
\end{figure}

Since \sysname sustained 6,000 TPS without any issue, we ran a second experiment with a load of 8,500 TPS. The performance of the same cluster using \sysname under this increased load is demonstrated in \Cref{fig:sui-throughput}, where both throughput and latency remain steady during five hours of sustained load.

As of January 2026, \sysname is deployed on Sui's devnet, and several testnet full nodes have been permanently migrated from RocksDB to \sysname.

\section{Related Work}\label{sec:related}

\paragraph{LSM-Tree Stores}
LSM-trees~\cite{oneil1996lsm} organize data into sorted runs across multiple levels,
trading write amplification for read performance through background compaction.
Stores like RocksDB~\cite{rocksdb2021evolution, cao2020rocksdb},
LevelDB~\cite{leveldb2011google}, and PebblesDB~\cite{raju2017pebblesdb}
have become industry standards, with various optimizations including
lazy leveling~\cite{dayan2018dostoevsky}, optimal Bloom filter allocation~\cite{dayan2017monkey},
and concurrent compaction~\cite{hyperleveldb2013hyperdex}.
Despite these improvements, compaction fundamentally rewrites data 5--10$\times$ across levels.
\sysname sidesteps this entirely: its append-only WAL eliminates multi-level compaction,
achieving 8.4$\times$ higher write throughput than RocksDB.

\paragraph{Key-Value Separation}
WiscKey~\cite{lu2016wisckey} pioneered storing values in a separate log (vLog),
dramatically reducing write amplification for large values.
Production systems including BlobDB~\cite{blobdb}, Titan~\cite{titan2019pingcap},
and DiffKV~\cite{li2021diffkv} build on this approach,
while HashKV~\cite{chan2018hashkv} and BVLSM~\cite{li2025bvlsm} optimize
value placement and garbage collection.
Kreon~\cite{kreon2021tucana} uses memory-mapped KV separation for flash storage,
while BadgerDB~\cite{badgerdb2017dgraph} implements WiscKey's design in Go.
These systems still maintain an LSM-tree for keys, inheriting compaction overhead.
\sysname differs fundamentally: the WAL serves as the unified value store
with lazy-flushed index tables, avoiding both LSM compaction and separate vLog management.

\paragraph{B-Tree and Memory-Mapped Stores}
LMDB~\cite{lmdb2014chu} uses copy-on-write B+ trees with shadow paging,
enabling zero-copy reads but limiting to single-writer concurrency.
CedrusDB~\cite{yin2020cedrusdb} combines memory-mapped lazy-tries with WAL,
the most architecturally similar prior work to \sysname.
Traditional systems like BerkeleyDB~\cite{berkeleydb1999olson} and
SQLite~\cite{sqlite2000hipp} use B-trees with logging and locking,
while redb~\cite{redb2023berner} offers a modern Rust implementation.
Unlike B-tree stores requiring in-place updates or copy-on-write,
\sysname's append-only design avoids page splits and tree rebalancing entirely.

\paragraph{Hardware-Optimized Stores}
KVell~\cite{lepers2019kvell} maximizes NVMe bandwidth via per-CPU slices
without data sorting, sharing \sysname's philosophy of avoiding sort overhead.
SpanDB~\cite{chen2021spandb} places WAL on fast NVMe with SPDK for parallel I/O,
while FASTER~\cite{chandramouli2018faster} uses epoch-protected hybrid logs
achieving 160M ops/sec.
SILK~\cite{balmau2019silk} and TRIAD~\cite{balmau2019triad} address LSM write stalls
through I/O scheduling and tiered storage.
\sysname takes a different approach: rather than optimizing around LSM limitations,
it eliminates them architecturally while using standard mmap and O\_DIRECT interfaces.

\iffullversion
\paragraph{Learned and Adaptive Indexes}
Machine learning techniques can replace traditional index structures with models
that predict key positions.
Bourbon~\cite{dai2020bourbon} adds learned indexes atop WiscKey using piecewise linear regression
for 1.23--1.78$\times$ faster lookups.
Kraska et al.~\cite{kraska2018learned} propose replacing B-trees and Bloom filters
with neural networks, trading training overhead for lookup speed.
ADOC~\cite{yu2023adoc} automatically tunes RocksDB parameters, achieving 87.9\% write stall reduction.
\sysname uses conventional Bloom filters and hash-based indexing for predictability,
prioritizing write performance over learned lookup optimizations.

\paragraph{Alternative Tree Structures}
SplinterDB~\cite{conway2020splinterdb} uses size-tiered B$\varepsilon$-trees,
achieving 6--10$\times$ faster insertions than RocksDB by batching updates in nodes.
TerarkDB~\cite{terarkdb2021bytedance} uses compressed searchable SSTs optimized for reduced tail latency.
ForestDB~\cite{forestdb2014couchbase} uses HB+-tries for document workloads,
while the Bw-Tree~\cite{levandoski2013bwtree} achieves lock-freedom through delta records
and epoch-based reclamation.
\sysname's append-only WAL avoids the complexity of tree maintenance entirely,
trading sorted iteration for minimal write amplification.

\paragraph{Rust Implementations}
Rust's memory safety guarantees make it increasingly popular for storage engines.
Fjall~\cite{fjall2025rs} provides modern LSM-trees with built-in KV separation.
AgateDB~\cite{agatedb2021tikv} ports BadgerDB's WiscKey-style design for TiKV.
sled~\cite{sled2019rs} uses flash-optimized Bw-trees with lock-free reads.
\sysname shares Rust's safety benefits while introducing a novel WAL-centric architecture
distinct from these LSM and B-tree based designs.

\paragraph{Blockchain Storage}
Blockchain systems require high-throughput persistent storage with strict durability.
\sysname was developed for Sui~\cite{sui2023mysticeti}, where transaction objects
(typically 1KB--1MB) drive the large-value workload.
Aptos uses JellyfishMerkle~\cite{aptos2022jellyfish} sparse Merkle trees for state management,
while Ethereum~\cite{ethereum2015wood} evolved from LevelDB to Pebble for state storage.
\sysname could serve as the underlying storage for such Merkle tree implementations.

\paragraph{Write-Ahead Logging}
ARIES~\cite{mohan1992aries} established foundational WAL protocols with
steal/no-force buffer management for B-tree databases.
Silo~\cite{tu2013silo} uses epoch-based transaction processing with group commit,
inspiring \sysname's batched WAL syncing.
Hekaton~\cite{diaconu2013hekaton} optimizes logging for in-memory OLTP,
while NVLogging~\cite{arulraj2018nvlogging} targets persistent memory.
Unlike traditional systems where WAL is auxiliary for recovery,
\sysname's WAL is the primary data store with lazy index flushing.
\fi

\section{Conclusion}

\sysname is a key-value store that eliminates value compaction by treating the WAL as permanent storage: values are written once and never moved, while small, lazily-flushed index tables map keys to WAL positions. On 1\,KB values, \sysname achieves 8.4$\times$ higher write throughput than RocksDB, 1.7$\times$ faster point queries, and 15.6$\times$ faster existence checks, with LSM-trees regaining advantage only below 128 bytes. Integration with Sui validators confirms these gains hold in production, maintaining stable performance under loads that collapse RocksDB-backed systems. The tradeoff is generality for efficiency on an increasingly common pattern: hash-keyed, kilobyte-scale, write-heavy workloads found in content-addressable storage, deduplication, and blockchain systems.

\section*{Acknowledgments}
This work is supported by Mysten Labs.
We sincerely thank John Martin for help in deploying \sysname within the Sui blockchain. We also thank Dmitri Perelman, Brandon Williams, Mark Logan, Inga Agibalova, and George Danezis for their support and feedback on early drafts.

\balance
\bibliographystyle{ACM-Reference-Format}
\bibliography{references}


\end{document}